\documentclass[twocolumn]{aastex631}

\usepackage{comment, amsmath}

\shorttitle{Pressure Estimates Around Star-Forming Regions}
\shortauthors{Murphy}

\begin{document}

\title{The Relative Importance of Thermal Gas, Radiation, and Magnetic Pressures Around Star-Forming Regions in Normal Galaxies and Dusty Starbursts}

\author[0000-0001-7089-7325]{Eric J.\,Murphy}
\email{emurphy@nrao.edu}
\affiliation{National Radio Astronomy Observatory\\ 
520 Edgemont Road\\
Charlottesville, VA 22903, USA}




\begin{abstract}
In this paper, an investigation on the relative importance of the thermal gas, radiation, and (minimum-energy) magnetic pressures around $\approx$200 star-forming regions in a sample of nearby normal and luminous infrared galaxies is presented. Given the range of galaxy distances, pressure estimates are made on spatial scales spanning $\sim$0.1$-3$\,kpc. The ratio of thermal gas-to-radiation pressures does not appear to significantly depend on star formation rate surface density ($\Sigma_{\rm SFR}$), but exhibits a steady decrease with increasing physical size of the aperture over which the quantities are measured. The ratio of magnetic-to-radiation pressures appears to be relatively flat as a function of $\Sigma_{\rm SFR}$ and similar in value for both nuclear and extranuclear regions, but unlike the ratio of thermal gas-to-radiation pressures, exhibits a steady increase with increasing aperture size. Furthermore, it seems that the magnetic pressure is typically weaker than the radiation pressure on sub-kpc scales, and only starts to play a significant role on few-kpc scales. When the internal pressure terms are summed, their ratio to the ($\Sigma_{\rm SFR}$-inferred) kpc-scale dynamical equilibrium pressure estimates is roughly constant. Consequently, it appears that the physical area of the galaxy disk, and not necessarily environment (e.g., nuclear vs. extranuclear regions) or star formation activity, may play the dominant role in determining which pressure term is most active around star-forming regions. These results are consistent with a scenario in which a combination of processes acting primarily on different physical scales work collectively to regulate the star formation process in galaxy disks.
\end{abstract}



\section{Introduction} \label{sec:intro}
Star formation is a slow, inefficient process that appears to be largely invariant with galaxy type or the physical scales it is averaged over \citep[e.g.,][]{ks98,rck07,kt07,bigiel08,akl08,lada12,utomo18}.  
Globally, observations suggest that galaxies are typically able to convert only a few percent of their available gas into stars per unit dynamical time, while within galaxies, it is found that $\lesssim$1\% of the available gas is able to be converted into stars per unit gravitational free-fall time.  
To date, a number of scenarios have been proposed to help explain the inefficiency of star formation, though a dominant process has yet to clearly emerge.  

Magnetic fields, cosmic-rays and/or externally-driven turbulence that can pervade the entire disks of galaxies \citep[e.g.,][]{parker69,sellwood99,ostriker01} have all been proposed as a means to prohibit gas from collapsing into newly-formed stars. 
The dynamical importance of magnetic fields and cosmic rays have also been demonstrated observationally by showing that their energy densities can be comparable to the thermal and turbulent gas energy densities on large scales \citep[e.g.,][]{beck07, fat08}.
However, stellar feedback processes that can operate on the scales of individual giant molecular clouds almost certainly play a substantial role driving the inefficiency of star formation \citep[e.g.][]{krumholz14}. 

Massive ($\gtrsim8\,M_{\odot}$) stars are able to inject large amounts of energy and momentum into the interstellar medium (ISM) of galaxies through numerous mechanisms.  
Models for feedback in the ISM have largely included energy and momentum injection by supernova, stellar winds, and radiation pressure on dust \citep[][]{mckee77,silk97,nzs03,tqm05,btd11,mqt10}.  
Furthermore, observations have additionally provided support for thermal gas and radiation pressure on dust being potentially important regulators of star formation both for individual massive star clusters \citep[e.g.,][]{nzs01, murray11}, as well as for entire starbust galaxies \citep[e.g.,][]{at11}.

In this paper, a combination of mid-infrared and multi-frequency radio data are used to investigate the relative strengths of thermal gas, radiation, and magnetic pressures on $\sim$100\,pc to $\sim$3\,kpc scales around star-forming regions within a sample of nearby galaxies included in the Star Formation in Radio Survey \citep[SFRS;][]{ejm12b, ejm18a} and luminous infrared galaxies (LIRGs; $L_{\rm IR} > 10^{11}\,L_{\sun}$) in the Great Observatories All-Sky LIRG Survey \citep[GOALS;][]{lee09}.  
The paper is organized as follows: 
The data and analysis procedures are presented in Section~\ref{sec:data}. 
The results are then presented in Section~\ref{sec:results}.  
A brief discussion of the findings and main conclusions are given in Section~\ref{sec:disc}.

\section{Data and Analysis}\label{sec:data}
To investigate the relative roles played by thermal gas, radiation and magnetic pressures within galaxy disks, a combination of mid-infrared and multi-frequency radio data are used for a sample of relatively nearby galaxies.  
The SFRS galaxy sample is derived from the {\it Spitzer} Infrared Nearby Galaxies Survey \citep[SINGS;][]{rck03} and Key Insights on Nearby Galaxies: a Far-Infrared Survey with {\it Herschel} \citep[KINGFISH;][]{kf11} legacy programs.  
Consequently, all sources are well studied and have a wealth of ancillary data available.  
Radio data were collected for a total of 112 star-forming complexes (50 nuclei and 62 extranuclear regions) easily observable with the VLA (i.e., having $\delta > -35\degr$).

The GOALS galaxy sample includes over 200 LIRGs included in the flux density-limited ($S_{\rm 60\,\mu m} > 5.24$\,Jy) IRAS Revised Bright Galaxy Sample \citep[RBGS;][]{rbgs03}.  
A subset of 68 GOALS galaxies within the declination range spanning $-20\degr < \delta < 20\degr$ were observed with the VLA as part of the GOALS ``equatorial" survey and included in the present study.  


\subsection{SFRS Data}
Radio data were obtained as part of multiple campaigns using the VLA covering the S- ($2-4$\,GHz), Ku- ($12-18$\,GHz), and Ka- ($26.5-40$\,GHz) bands.  
Details on the data reduction and imaging procedures can be found in \citet{ejm18a,stl20}.  
To summarize, standard calibration procedures are followed, using the VLA calibration pipeline built on the Common Astronomy Software Applications \citep[CASA;][]{casa} versions 4.6.0 and 4.7.0.
After each initial pipeline run, the calibration tables and visibilities are manually inspected and flagged for signs of instrumental problems (e.g., bad deformatters) and radio frequency interference (RFI). 
After flagging, the pipeline is re-run, with this process repeated until any further signs of bad data could not be detected.
As with the data calibration, a detailed description of the imaging procedure used here can be found in \citet{ejm18a, stl20}.  
Calibrated VLA measurement sets for each source were then imaged using the task \textsc{tclean} in CASA version 4.7.0.

We also make use of archival infrared and UV data.  
{\it Spitzer} 24\,$\mu$m data were taken from the SINGS and Local Volume Legacy (LVL) legacy programs.
Details on the associated observation strategies and data reduction steps can be found in \citet{dd07} and \citet{dd09}, respectively.  
Two galaxies, IC\,342 and NGC\,2146, were not a part of SINGS or LVL; their 24\,$\mu$m imaging comes from \citet{ce08}.  
{\it GALEX} far-UV (FUV; 1528Å) data were taken from the {\it GALEX} Large Galaxy Atlas \citep{seibert07}. 

\subsection{GOALS Data}
Similar to the SFRS observations, radio data for the GOALS  equatorial sample were obtained using the VLA covering the S- ($2-4$\,GHz), Ku- ($12-18$\,GHz), and Ka- ($26.5-40$\,GHz) bands.  
Calibration and imaging procedures are identical to those described above for the SFRS radio data.  
More details on the data reduction and imaging procedures, which follows those described in \citet{ejm18a}, can be found in \citet{stl19}.  

Archival {\it Spitzer} 8\,$\mu$m data taken as part of GOALS are also used in this analysis as the basis for total infrared luminosity ($L_{\rm 8-1000\,\mu m}$) estimates of each region. 
Details on the associated observation strategies and data reduction steps are available in J. M. Mazzarella et al. (2022, in preparation).  
The conversion between the 8\,$\mu$m and total infrared luminosities rely on the study of \citet{de11} and assume a fixed ratio  ${\rm IR8} = L_{\rm IR}/L_{\rm 8\mu m} = 8.1\pm2$ \citep[see][for more details]{stl19}.

\begin{figure*}
\epsscale{1.15}
    \centering
    \plottwo{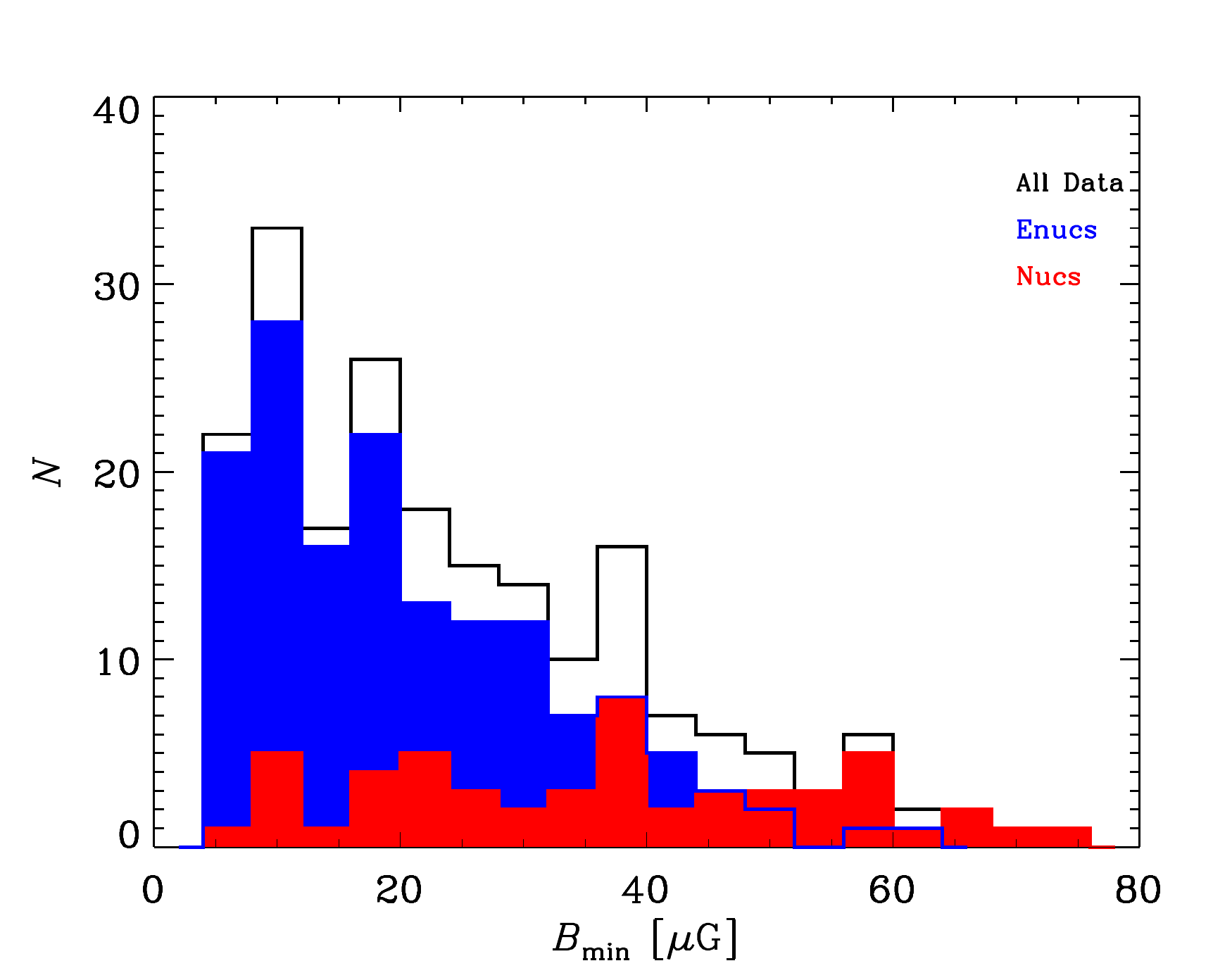}{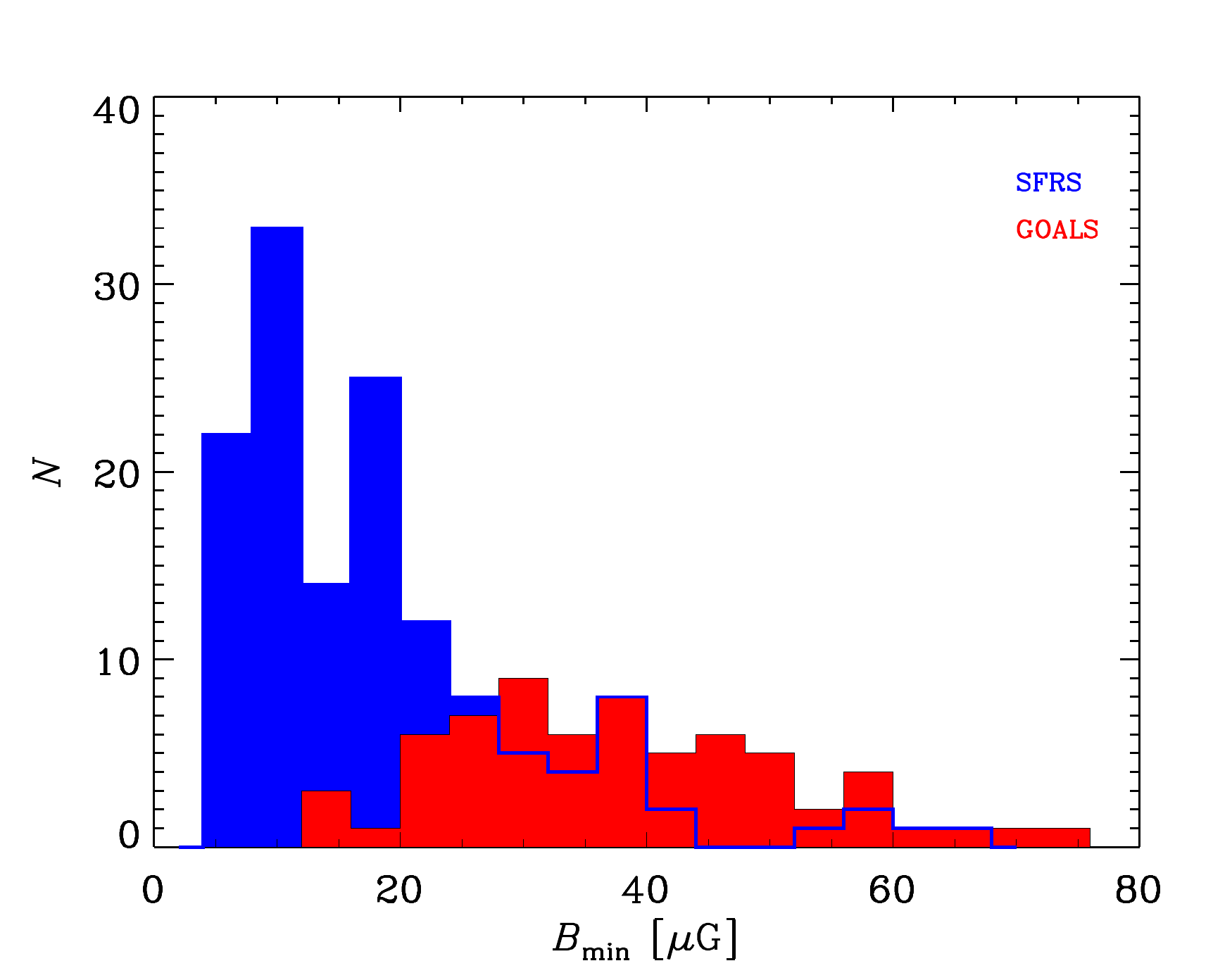}
    \caption{{\it Left:} Minimum-energy magnetic field strength distribution for all sources, along with the distributions for nuclear and extranuclear regions. 
    The field strengths for the extranuclear regions are, on average, smaller and more narrowly distributed than those for the nuclear regions.  
    {\it Right:} Minimum-energy magnetic field strength distributions for regions within  the SFRS and GOALS galaxies.  
    The field strengths of regions within the SFRS galaxies are, on average, smaller and more narrowly distributed than those within the GOALS sources. }
    \label{fig:Bmindist}
\end{figure*}

\begin{figure*}
\epsscale{1.15}
    \centering
    \plottwo{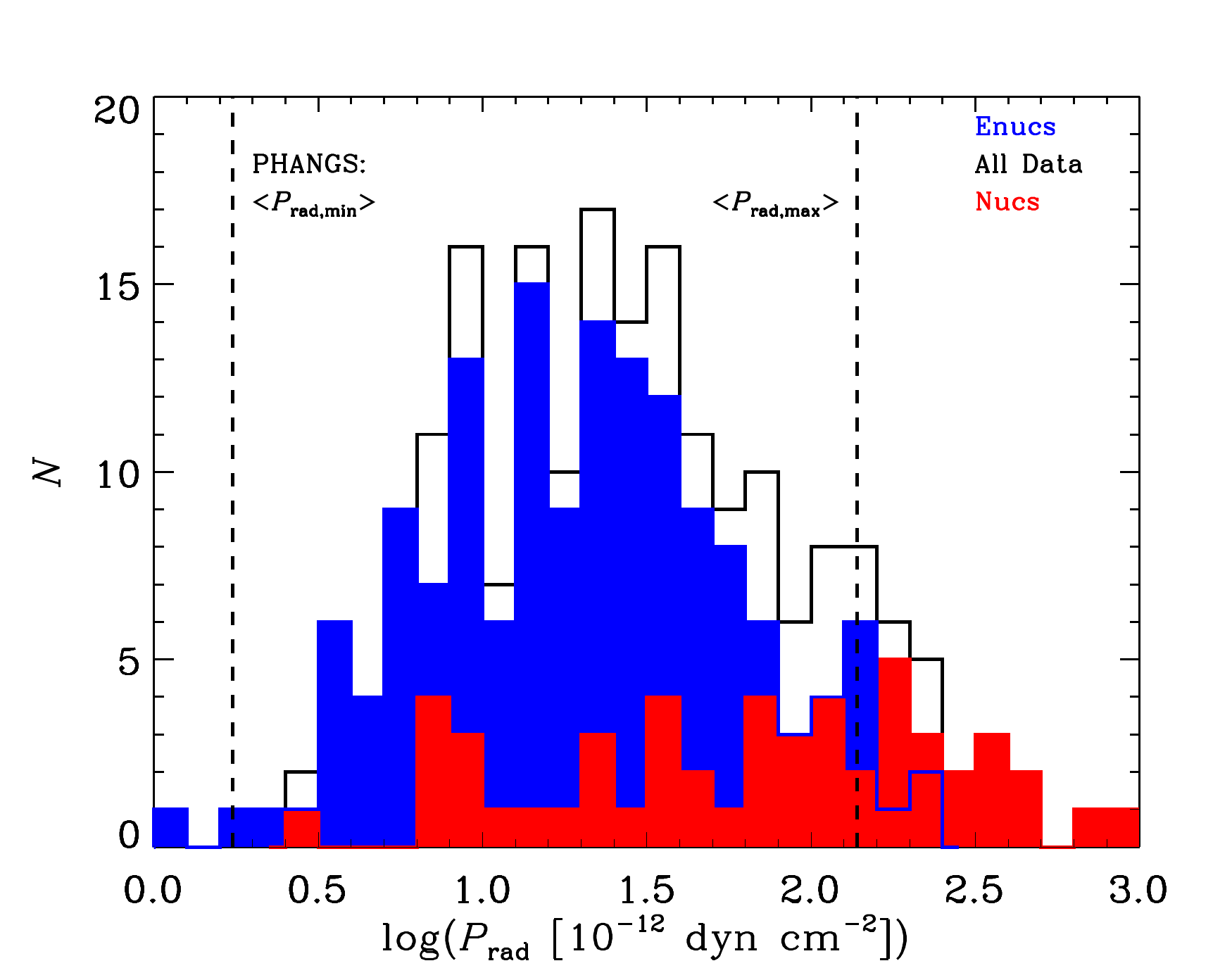}{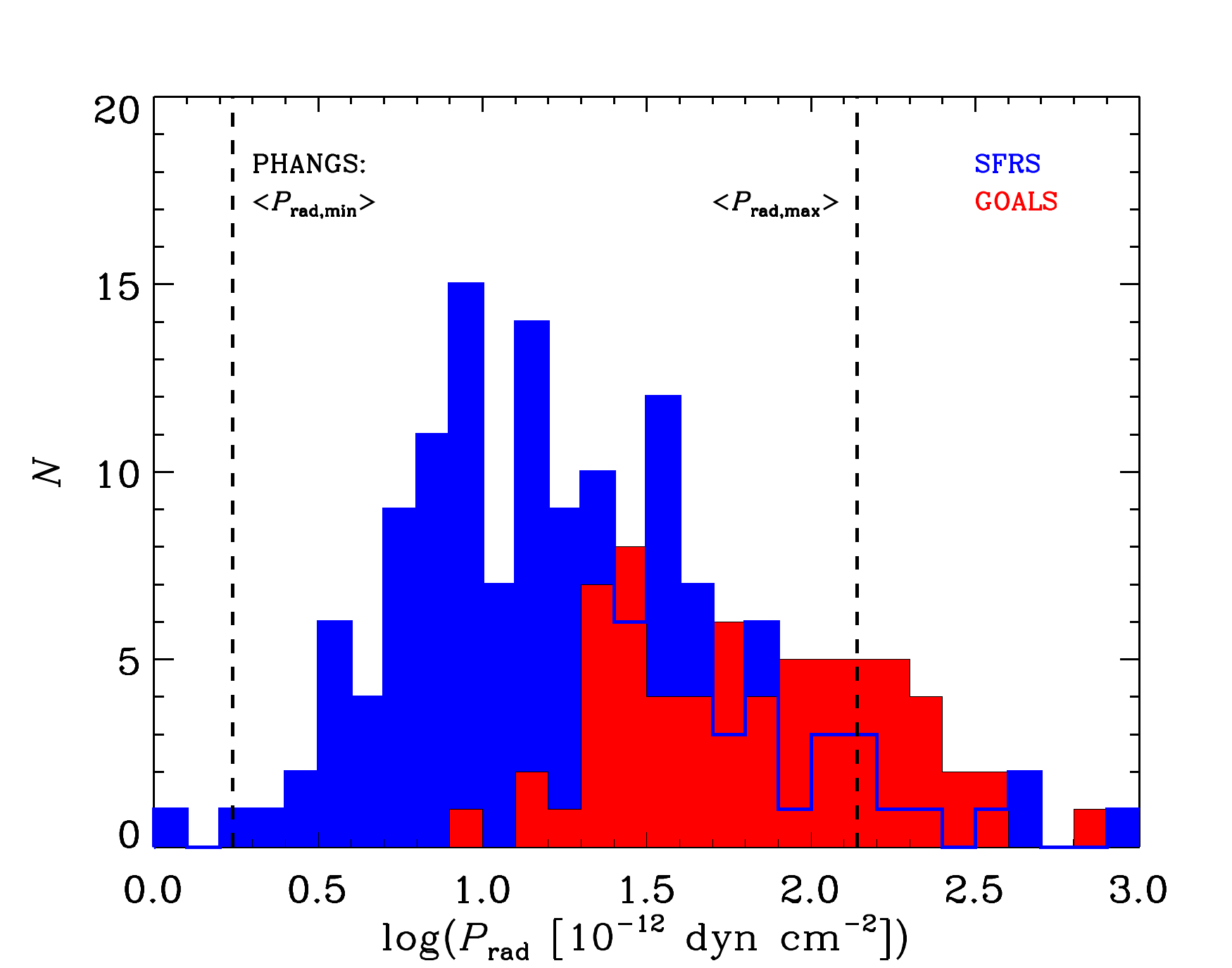}
    \caption{{\it Left:} Radiation pressure distribution for all sources, along with the distributions for nuclear and extranuclear regions. 
    Similar to the minimum-energy magnetic field strength distribution, the radiation pressure estimates for the extranuclear regions are, on average, smaller and more narrowly distributed than those for the nuclear regions.  
    {\it Right:} Radiation pressure distributions for regions within the SFRS and GOALS galaxies.  
    The radiation pressures for regions within the SFRS galaxies are, on average, smaller than those within the GOALS sources. 
    The vertical lines in both figures indicate the minimum and maximum radiation pressure estimates averaged over a sample of H{\sc ii} regions measured on ${\sim}50{-}100$\,pc scales by \citet{barnes21}, indicating that our measurements largely fall between their extreme average values.}
    \label{fig:Praddist}
\end{figure*}

\begin{figure*}
\epsscale{1.15}
    \centering
    \plottwo{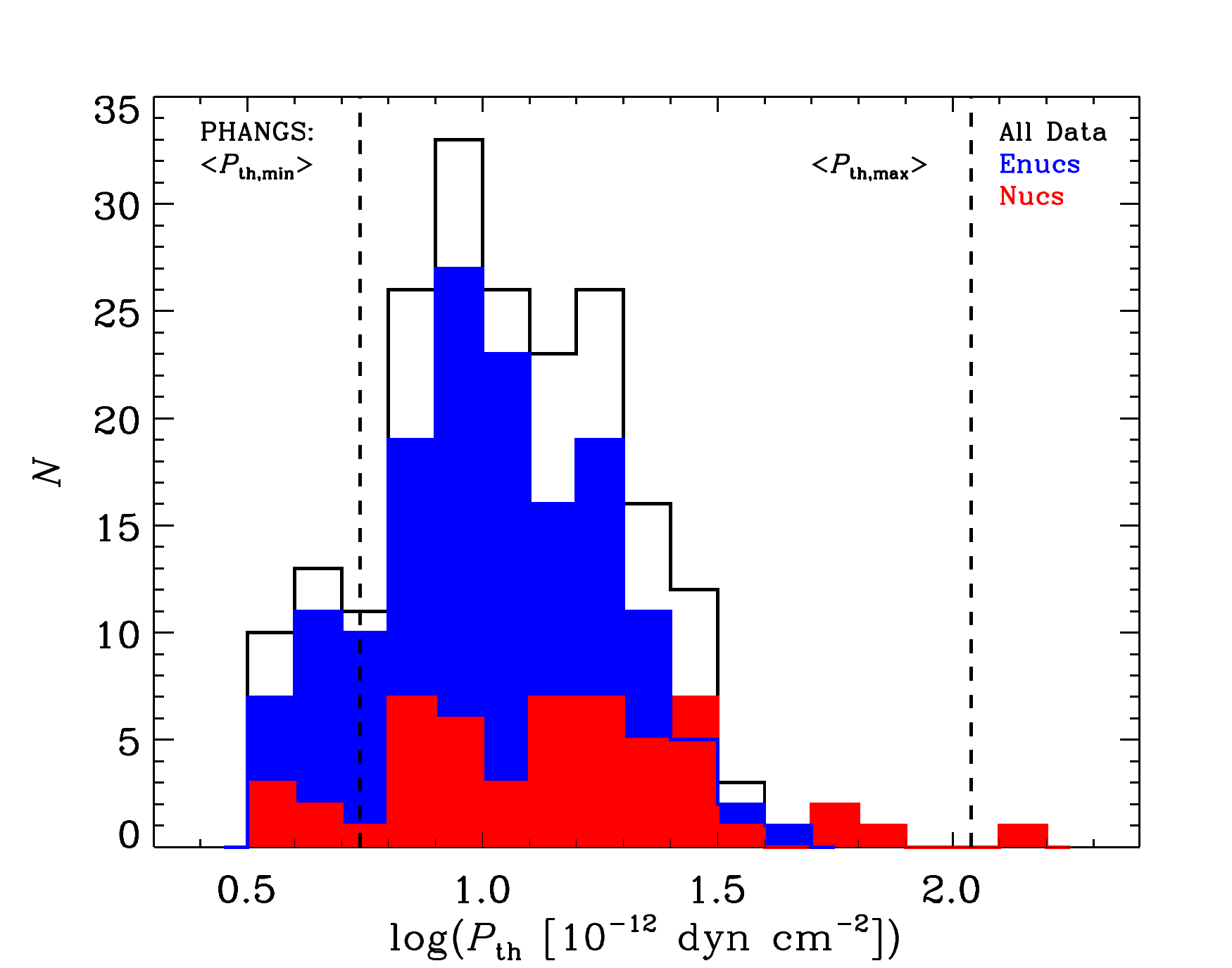}{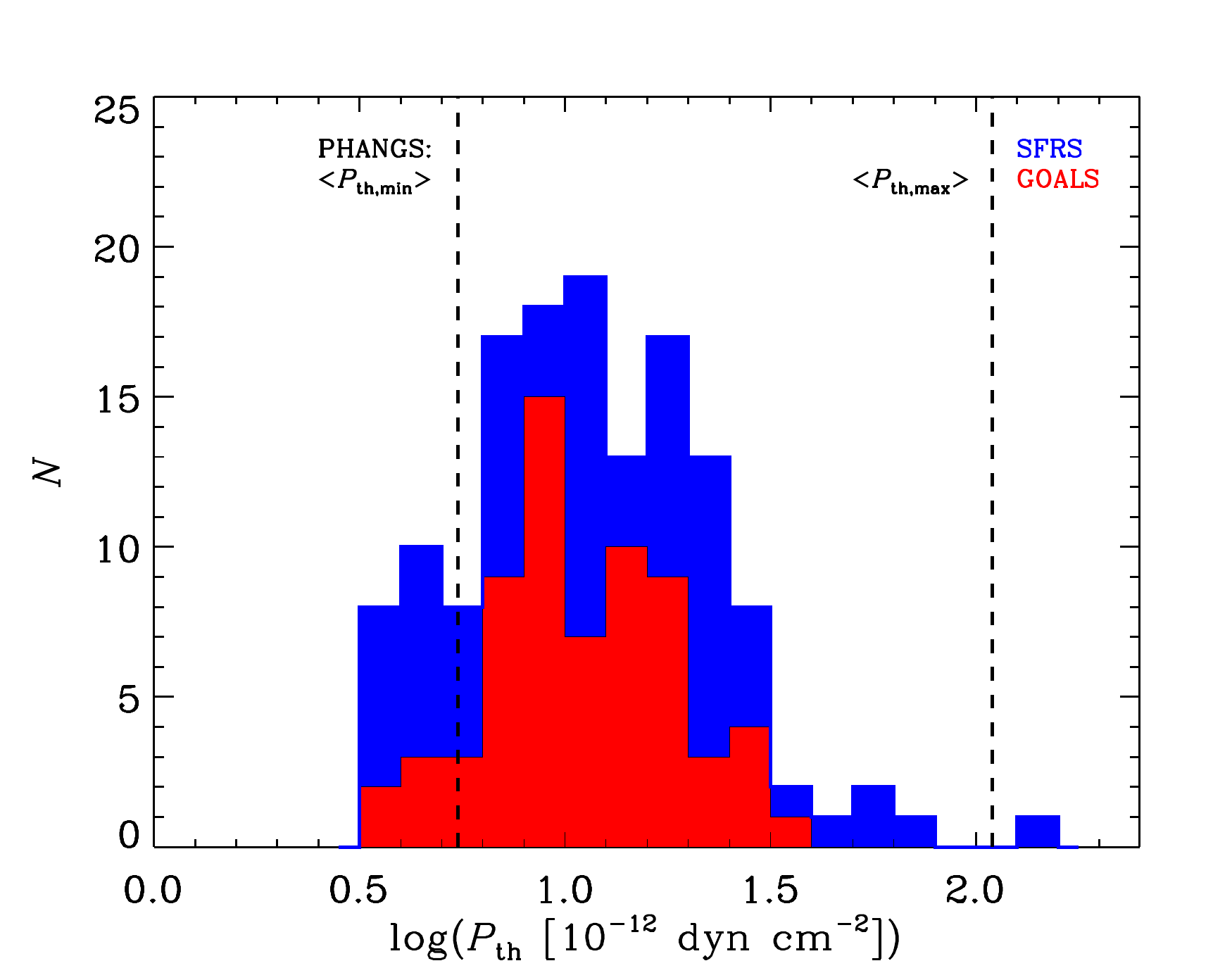}
    \caption{{\it Left:} Thermal gas pressure distribution for all sources, along with the distributions for nuclear and extranuclear regions. 
    Unlike what is seen for the radiation pressure distributions, the thermal gas pressure estimates for the nuclear and extranculear regions are, on average, quite similar.   
    {\it Right:} Thermal gas pressure distributions for regions within the SFRS and GOALS galaxies.  
    The thermal gas pressure distributions for regions within the SFRS and GOALS galaxies are quite similar.   
    The vertical lines in both figures indicate the minimum and maximum thermal gas pressure estimates averaged over a sample of H{\sc ii} regions measured on ${\sim}50{-}100$\,pc scales by \citep{barnes21}, indicating that our measurements largely fall between their extreme average values.}
    \label{fig:Pthdist}
\end{figure*}

\subsection{Photometry}
Details on the SFRS photometry can be found in \citet{stl20}.  
To summarize, photometry on the full VLA and {\it Spitzer}/MIPS 24\,$\mu$m data sets were carried out after matching their resolution (7\arcsec~FWHM), cropping each image to a common field-of-view, and regridding to a common pixel scale.  
Apertures having a diameter of 7\arcsec~were used to measure the emission from each region within the galaxies.  
Given the distance of the SFRS galaxies, this angular diameter corresponds to linear values ranging from $\sim$100$-1000$\,pc, with a median of $\sim$260\,pc.   
Flux densities and associated uncertainties for all star-forming regions used in the present analysis are given in \citet{stl20}.  
Sources identified as likely background galaxies, possibly being associated with supernovae, or potential anomalous microwave emission candidates in \citet{stl20}, are not included in this study.

Details on the GOALS photometry can be found in \citet{stl19}.  
Similar to the handling of the SFRS data, photometry on the GOALS radio and 8\,$\mu$m imaging was carried out after first matching their resolution (2\farcs5~FWHM), cropping each image to a common field-of-view, and regridding to a common pixel scale.  
Apertures having a diameter of 4\arcsec~were used to measure the emission from each region with the galaxies. 
Given the distance of the GOALS galaxies, this angular diameter corresponds to linear values ranging from $\sim$1$-2.8$\,kpc, with a median of $\sim$1.7\,kpc.   
Flux densities and associated uncertainties for all regions are given in \citet{stl19}.

Given that the photometry comes from two catalogs having used photometric apertures of differing sizes relative to the angular resolution of the maps, aperture corrections are applied assuming that the source emission profiles are largely Gaussian.  
The SFRS photometry is therefore multiplied by a factor of 1.72 and the GOALS photometry is multiplied by a factor of 1.13.  The final sample includes a total of 204 sources: 31 nuclear (i.e., those at a galactocentric radius $r_{\rm g} < 250$\,pc) and 107 extranuclear star forming regions from the SFRS sample, along with 22 nuclear and 44 extranuclear star-forming regions from the GOALS sample.

\subsection{Separation of Non-Thermal and Free-Free Emission}\label{sec:tfrac}
Using the multi-frequency radio data, the non-thermal and free-free emission is separated to independently estimate the minimum-energy magnetic field strengths and star formation rates, as described below.  
The method used to separate the non-thermal and free-free components is described in detail in \cite{ejm18a,stl19,stl20}.  
However, rather than measuring the relative fractions at 33\,GHz, measurements at 3\,GHz are used as this frequency is more sensitive to the amount of non-thermal emission necessary to estimate the minimum-energy magnetic field strengths.

Briefly, the fraction of free-free emission at 3\,GHz was estimated using he prescription in \citet{kwb84}, such that

\begin{equation} 
\label{eq:tfrac}
f_{\rm T}^{\nu_{1}} = \frac{(\frac{\nu_{2}}{\nu_{1}})^{\alpha} - (\frac{\nu_{2}}{\nu_{1}})^{\alpha^{\rm NT}}}{(\frac{\nu_{2}}{\nu_{1}})^{-0.1} - (\frac{\nu_{2}}{\nu_{1}})^{\alpha^{\rm NT}}}
\end{equation}

\noindent
where $\nu_{1}$ is the target frequency (3\,GHz), $\alpha$ is the observed spectral index between 3 and 33 GHz, $\alpha^{\rm NT}$ is the non-thermal spectral index, and a single power-law exponent for the free-free emission ($\sim -0.1$) is assumed.
The non-thermal spectral index is fixed at $\alpha^{\rm NT} = -0.83$ except when $\alpha < -0.83$ or $\alpha_{3-15\,{\rm GHz}} < -0.83$, in which case the lesser (steeper) values are used to set the lower limit on $\alpha^{\rm NT}$.    
A constant non-thermal radio spectral index of $-0.83$ is assumed based on the average non-thermal spectral index found among the 10 star-forming regions studied in NGC\,6946 by \citet{ejm11b}. 
This value is additionally consistent with the results of \citet[][i.e., \(\alpha^{\rm NT} = -0.83\) with a scatter of \(\sigma_{\alpha^{\rm NT}} = 0.13\)]{nb97} for a sample of 74 nearby galaxies, although quite a bit steeper than the mean value of $-0.6$ measured by \citet{klein18} for a sample of 14 galaxies when employing more sophisticated models for radio spectra than simple power laws.

While the calculated thermal fractions are clearly sensitive to the non-thermal spectral index, \citet{stl20} showed via a Monte Carlo Markov Chain (MCMC) analysis that adopting a fixed value does not introduce any systematic biases over a reasonable set of physically-motivated input conditions for a simple two-component power-law model for radio spectra.    
In order to test the sensitivity of the assumed non-thermal spectral index of $-0.83$ on the present analysis, the significantly flatter value of $\alpha_{\rm NT} = -0.6$ was instead assumed for all sources.  
This resulted in a median 3\,GHz thermal fraction of 0.37, which is only 20\% lower than the value when assuming \(\alpha_{\rm NT} = -0.83\) (i.e., 0.47), and thus would not impact the main conclusions of this paper. 

\subsection{Magnetic Pressure Estimates}
\label{sec:PB}
Taking the non-thermal flux densities, the revised equipartition and minimum-energy formulas of \citet{bk05} are used to calculate the minimum-energy magnetic field strength for each region. 
This calculation assumes a proton-to-electron number density ratio for particles in the energy range corresponding to GHz synchrotron emission of 100, a fixed non-thermal spectral index (as assumed above for the estimate of the thermal fractions), and a path length through the emitting medium equal to 1\,kpc.   
The distribution of minimum-energy magnetic fields for all regions are shown in Figure~\ref{fig:Bmindist}.


The values used for both the non-thermal spectral index and proton-to-electron number density ratio do not strongly affect the resulting minimum-energy magnetic field strength estimates.  
For example, choosing the rather extreme value of $\alpha_{\rm NT} = -0.6$ for all regions only increases the median minimum-energy magnetic field strength by $\approx$10\% (i.e., from 21.4 to 23.6\,$\mu$G), while doubling the proton-to-electron number density ratio to 200 for all regions, in the case of extreme energy losses to CR electrons, only increases the median minimum-energy magnetic field strength by $\approx$20\% (i.e., from 21.4 to 25.6\,$\mu$G).  
However, even for extreme starbursts where where hadronic losses may dominate, the assumption of a proton-to-electron number density ratio of $\sim$100 and the usage of the equipartition and minimum-energy formulas are expected to be valid \citep{lb13}.  

The choice of 1\,kpc for the path length should be appropriate for most regions, given that this is the value for the vertical synchrotron scaleheight in typical galaxy disks \citep[e.g.,][]{krause18}.  
This is also likely appropriate for nuclear regions that may drive outflows on similar scales, though such path lengths could in fact be larger.  
Where this assumption may not be valid is for H{\sc ii} regions resolved on $\sim$100\,pc scales, where shocks are able to drive enhancements to the local magnetic field, resulting in synchrotron emission in excess of the local medium.  
In such cases, smaller values of the path length may be more appropriate, which would increase the minimum-energy magnetic field strength estimates.  
However, as shown later in \S\ref{sec:results-int}, scaling the path length with the physical diameter for those extranuclear star-forming regions resolved at $<$1\,kpc scales after removing a local background does not qualitatively affect the results. 

For a relativistic gas the magnetic pressure, $P_{\rm B}$, is related to the magnetic field energy density such that
\begin{equation}
\label{eq:PB}
    P_{\rm B} = u_{\rm B} = \frac{B^2}{8\pi}, 
\end{equation} 
where, in cgs units, $P_{\rm B}$ is in units of ${\rm dyn\,cm^{-2}}$, $u_{\rm B}$ is in units of ${\rm erg\,cm^{-3}}$, and $B$ is in units of Gauss.  

Under the same minimum-energy assumptions, the cosmic-ray energy density is related to the magnetic field energy density, such that 
\begin{equation}
    u_{\rm CR} = \left(\frac{2}{1 - \alpha^{\rm NT}}\right) u_{\rm B}  
\end{equation}
\citep{bk05}, and the cosmic-ray pressure of the relativistic gas is related to the cosmic-ray energy density such that
\begin{equation}
\label{eq:PCR}
    P_{\rm CR} = \frac{u_{\rm CR}}{3}. 
\end{equation}
Assuming that the true magnetic field strength is near the minimum-energy value, the cosmic-ray pressure will be sub-dominant relative to the magnetic pressure.  
If this were not the case, Parker instabilities would occur, inflating bubbles that expel cosmic rays from the galaxy disks \citep{parker65}.


\subsection{Radiation Pressure Estimates}
To get an estimate for the radiation pressure of each star-forming region, an estimate of the bolometric surface brightness is required.  
For the SFRS galaxies, the bolometric surface brightness is estimated from each region by first calculating observed (i.e., non-extinction corrected) far-UV and infrared based star formation rates using the $GALEX$ FUV and {\it Spitzer} 24\,$\mu$m photometry along with Equations 2 and 14 in \citet{ejm12b}, respectively, where
\begin{equation}
\Biggl(\frac{\rm \psi_{FUV}}{M_{\sun}\,{\rm yr^{-1}}}\Biggr) = 4.42\times10^{-44}\Biggl(\frac{L_{\rm FUV}}{\rm erg~s^{-1}}\Biggr)
\end{equation}
and
\begin{equation}
\Biggl(\frac{\rm \psi_{24\,\micron}}{M_{\sun}\,{\rm yr^{-1}}}\Biggr) = 2.45\times10^{-43} \Biggl[\frac{\nu L_{\nu}(24\,\micron)} {\rm erg~s^{-1}} \Biggr].  
\end{equation}
These star formation rates are then summed and converted back into an estimate of the bolometric surface brightness by first estimating the bolometric flux, $F_{\rm bol}$, using Equation 15 in \citet{ejm12b} and then dividing by the area of the region, such that
\begin{equation}
\label{eq:Ibol}
I_{\rm bol} = \frac{F_{\rm bol}}{\Omega} \sim \Biggl(\frac{2.65\times10^{-7}}{\Omega d_{\rm L}^{2}}\Biggr) \Biggl(\frac{\psi_{\rm FUV} + \psi_{24\,\mu{\rm m}}}{M_{\sun}~{\rm yr^{-1}}}\Biggr),
\end{equation}
where $\Omega$ is in units of steradians, $d_{\rm L}$ is the luminosity distance to the source in Mpc, and $I_{\rm bol}$ is in cgs units of ${\rm erg\,s^{-1}\,cm^{-2}\,sr^{-1}}$. 
For the GOALS galaxies, the {\it Spitzer} 8\,$\mu$m-based total infrared luminosities given in \citet{stl19} are used as an estimate for the bolometric fluxes for these extremely dusty systems.  
With an estimate of the bolometric surface brightness for each region, the radiation pressure, $P_{\rm rad}$ is related to the radiation energy density for radiation emitted near the surface of a semitransparent body such that 
\begin{equation}
\label{eq:Prad}
    P_{\rm rad} = u_{\rm rad} \approx \left(\frac{2\pi}{c}\right)I_{\rm bol}. 
\end{equation} 
The distribution of radiation pressures for all regions are show in Figure~\ref{fig:Praddist}.

\subsection{Thermal Gas Pressure Estimates}
At radio frequencies, where $\tau \ll 1$, the ionizing photon rate is directly proportional to the free-free spectral luminosity, $L_{\nu}^{\rm ff}$, varying only weakly with electron temperature $T_{\rm e}$ \citep{rr68}, such that
\begin{equation}
\label{eq-Qo}
\begin{split}
\biggl[\frac{Q(H^{0})}{\rm s^{-1}}\Biggr] &= 6.3\times10^{25} \\ 
&\Biggl(\frac{T_{\rm e}}{10^{4}~{\rm K}}\Biggr)^{-0.45} \Biggl(\frac{\nu}{\rm GHz}\Biggr)^{0.1} \Biggl(\frac{L_{\nu}^{\rm T}}{\rm erg~s^{-1}~Hz^{-1}}\Biggr),  
\end{split}
\end{equation}
where an electron temperature of $T_{\rm e} = 10^{4}$\,K is assumed.  

Assuming that each region being investigated can be roughly approximated by an ionization bounded H{\sc ii} region in equilibrium, the ionizing photon rates can be used to estimate the corresponding electron densities such that, 
\begin{equation}   
    n_{\rm e} \gtrsim \sqrt\frac{3Q(H^{0})}{4\pi\alpha_{\rm H}R_{\rm s}^3}, 
\end{equation}
where, in cgs units, $n_{\rm e}$ is in units of ${\rm cm^{-3}}$, $R_{s}$ is Str\"{o}mgren radius in units of cm, taken as half of the photometric aperture diameter, and $\alpha_{\rm H} \approx 3\times10^{-13}\,{\rm cm^3\,s^{-1}}$ is the case B recombination coefficient for hydrogen.  
The $\gtrsim$ sign in the above equation comes from the fact that the photometric apertures are almost certainly upper limits to the actual H{\sc ii} region sizes.  

With an estimate for the electron density, the thermal gas pressure can also be estimated such that,
\begin{equation}
\label{eq:Pth}
P_{\rm th} = 2\times10^4 n_{\rm e} k_{\rm B} \Biggl(\frac{T_{\rm e}}{\rm 10^4\,K}\Biggl)
\end{equation}
where, in cgs units, $P_{\rm th}$ is in units of ${\rm dyn\,cm^{-2}}$ and $k_{\rm B} = 1.38\times10^{-16}\,{\rm cm^2\,g\,s^{-2}\,K^{-1}}$ is Boltzmann's constant.  
Again, an electron temperature of $T_{\rm e} = 10^{4}$\,K is assumed.
The factor of 2 is included under the additional assumption that hydrogen is fully ionized and helium is singly ionized, such that $n_{\rm e} + n_{\rm H} + n_{\rm He} \approx 2n_{\rm e} $.  
The distribution of thermal gas pressures for all regions are shown in Figure~\ref{fig:Pthdist}.

\subsection{Star Formation Rate Estimates}
Being directly proportional to the ionizing photon rate, optically-thin free-free emission can additionally be used to estimate star formation rates of each region using 
Equation 11 in \citet{ejm11b} such that
\begin{equation}
\label{eq:sfrt}
\begin{split}
\Biggl(\frac{\psi_{\rm ff}}{M_\odot \mathrm{\,yr}^{-1}}\Biggr) &= 4.6 \times 10^{-28}\\
&\Biggl(\frac{T_{\rm e}}{10^{4}~{\rm K}}\Biggr)^{-0.45} \Biggl(\frac{\nu}{\rm GHz}\Biggr)^{0.1}
\Biggl( \frac {L_{\nu}^{\rm ff}}{\mathrm{erg\,s^{-1}\,Hz}^{-1}} \Biggr).     
\end{split}
\end{equation}
Star formation rates are converted into star formation rate surface densities ($\Sigma_{\rm SFR}$) by dividing SFRs by the physical area of each aperture.  
By using the free-free emission, this estimate of the star formation rate is likely more robust than the infrared-based estimates as it does not depend on the fraction of photons that are reprocessed by dust relative to those that are able to escape each star-forming region or the fraction of dust that is heated by older stellar populations.  
Even so, the free-free to infrared-based star formation rates have a median ratio of unity with a median absolute deviation of $\approx$0.65.
It is likely worth noting that if a significantly flatter value for the non-thermal spectral index (i.e., $-0.6$) was assumed when carrying out the thermal/non-thermal decomposition of the radio emission as described in \S\ref{sec:tfrac}, the free-free and infrared based star formation rates become increasingly discrepant, decreasing from a median ratio of unity to a value of 0.89.


\begin{figure}
\epsscale{1.25}
    \centering
    \plotone{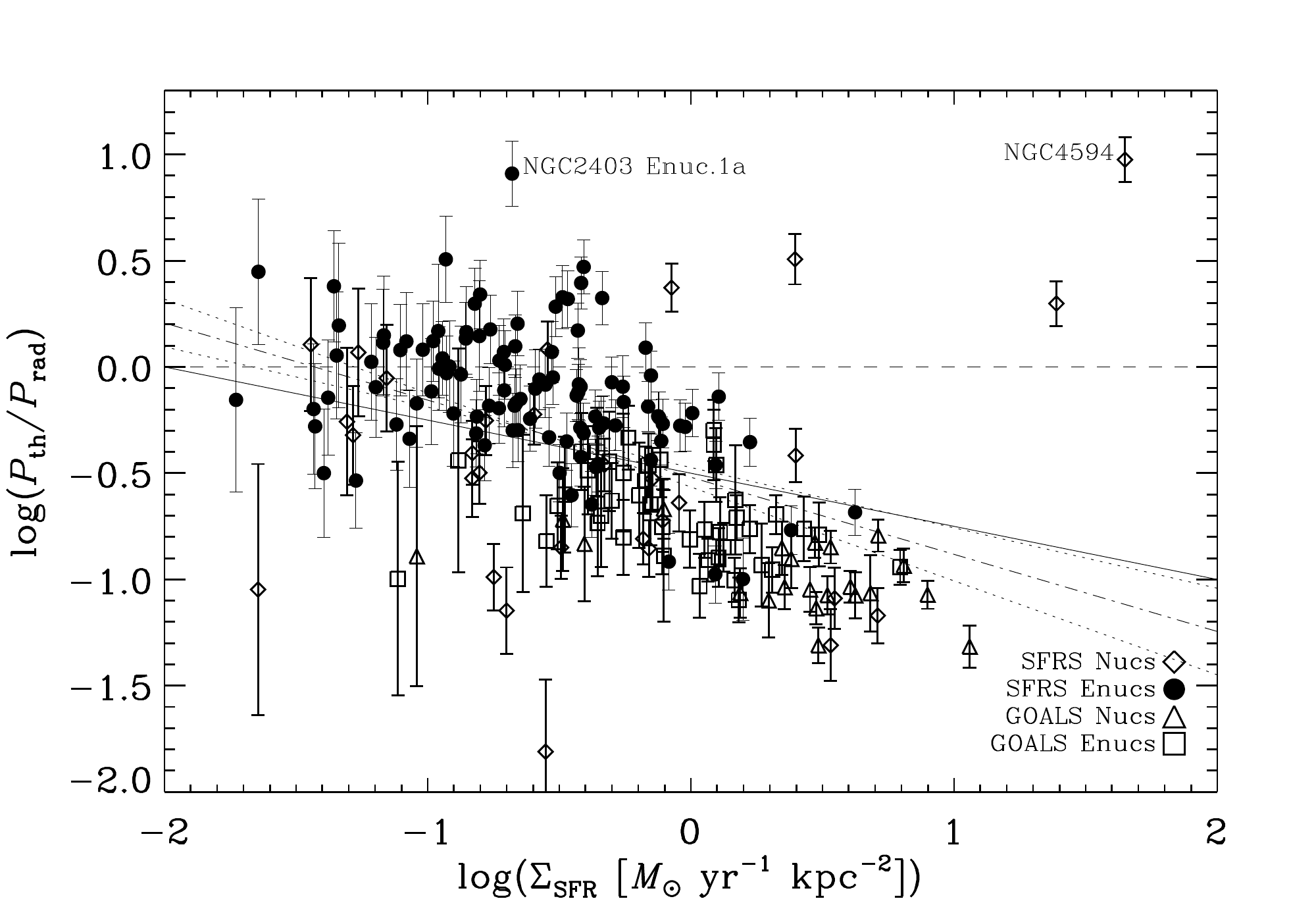}
    \caption{Logarithmic thermal gas-to-radiation pressure ratios plotted as a function of star formation rate surface density.  
    Subsets of the sample are identified using different plotting symbols.  
    The dashed line indicates a ratio of unity.  
    The thin solid line illustrates the expected trend given the correlated nature of the axes, as they both rely on the photometric aperture diameters. 
    The two sources having thermal gas-to-radiation pressure ratios $\gtrsim$5 are labeled: the nucleus of NGC\,4594 hosts an AGN \citep{jm10}; NGC\,2403\,Enuc.\,1\,A is likely an optically-faint SNR \citep{th94,sp03}.  
    A least-squares fit to the data is shown by a dot-dashed line (along with 1$\sigma$ uncertainties by the dotted-lines) indicating that the decrease in thermal gas-to-radiation pressure ratios with increasing values of $\Sigma_{\rm SFR}$ is only marginally steeper than what is expected given the correlated nature of the axes.  
        }   
    \label{fig:urat-th-sigsfr}
\end{figure}

\begin{figure}
\epsscale{1.25}
    \centering
    \plotone{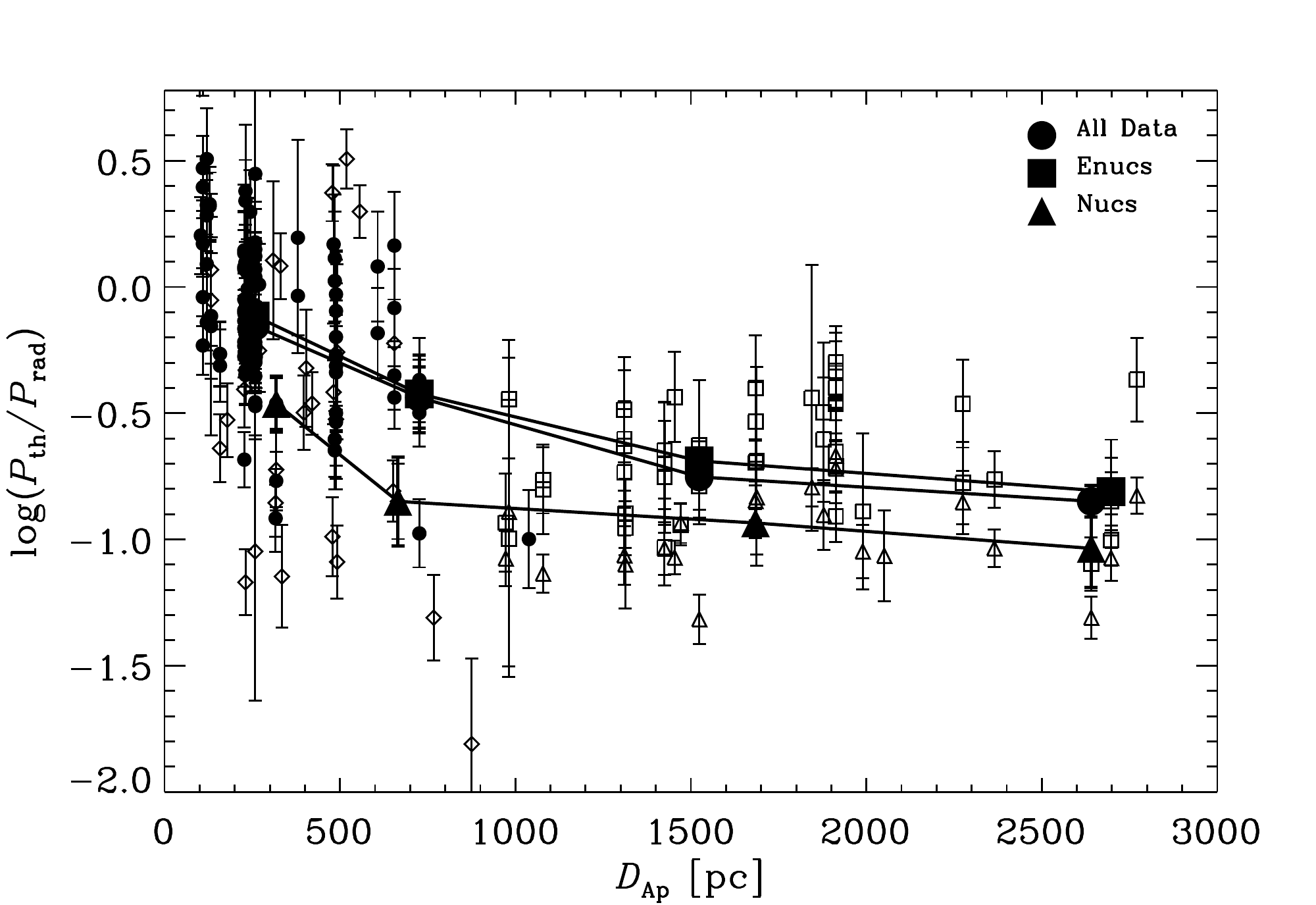}
    \caption{Logarithmic thermal gas-to-radiation pressure ratio plotted as a function of aperture diameter in physical units.   
    Subsets of the sample are identified using different plotting symbols: SFRS nuclear regions (open diamonds); SFRS extranuclear regions (filled circles); GOALS nuclear regions (open triangles); and GOALS extranuclear regions (open squares).  
    Running medians for all regions, as well as all nuclear and extranuclear regions, are shown by solid lines using large filled circles, triangles, and squares, respectively.  
    The thermal gas-to-radiation pressure ratios appear to decrease with increasing physical size of each region, indicating that the thermal pressure (i.e., free-free emission) is more centrally peaked on individual star-forming regions than the radiation pressure (i.e., infrared emission).  }
    \label{fig:apD-th}
\end{figure}

\begin{figure}
\epsscale{1.25}
    \centering
    \plotone{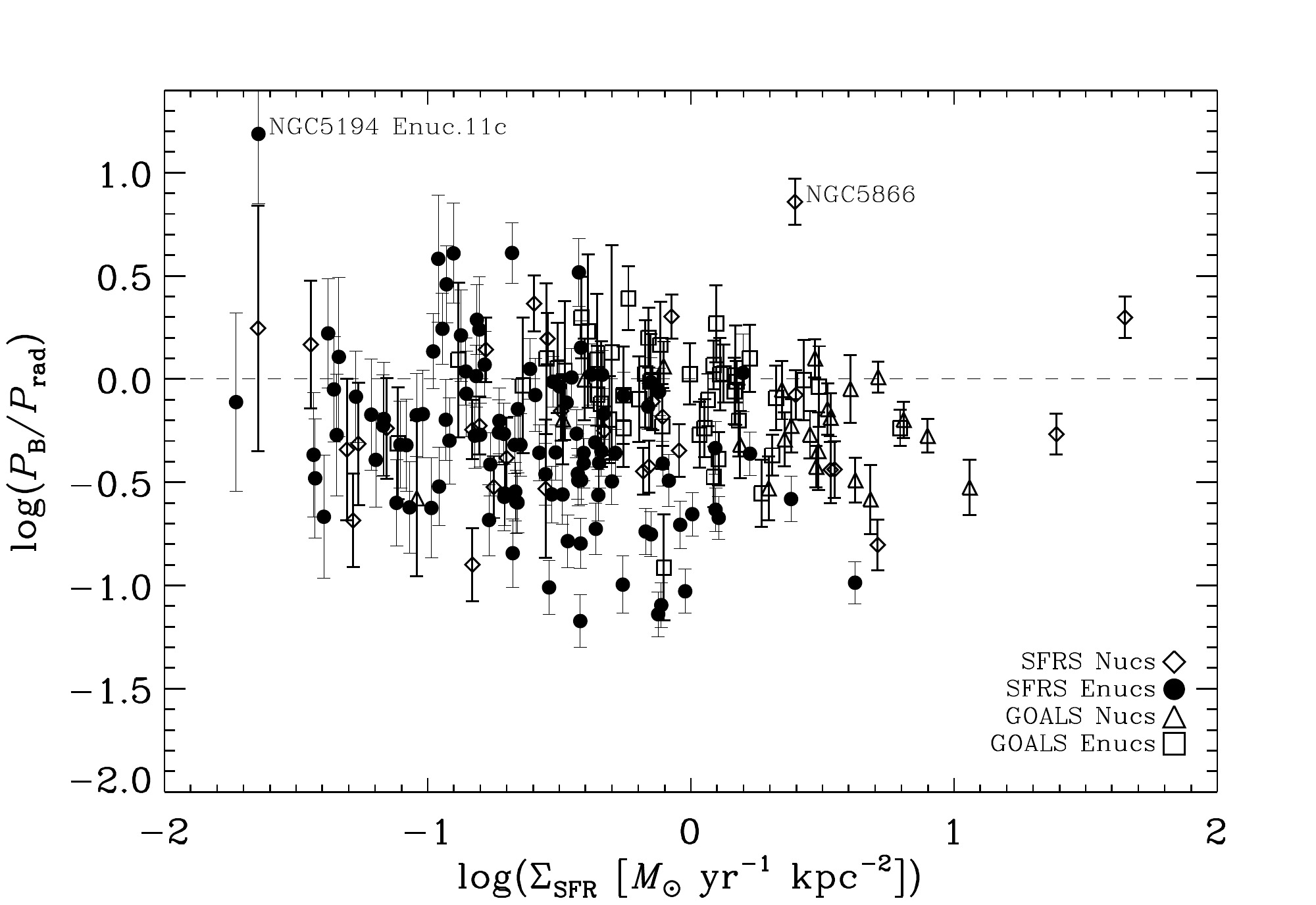}
    \caption{Logarithmic magnetic-to-radiation pressure ratios plotted as a function of star formation rate surface density.  
    Subsets of the sample are identified using different plotting symbols.  
    The dashed line indicates a ratio of unity.  
    The two sources having magnetic-to-radiation pressure ratios $\gtrsim$5 are labeled: the nucleus of NGC\,5866 hosts a known AGN \citep{jm10}; NGC\,5194\,Enuc.\,11\,C may be an optically-faint SNR \citep{maddox07}.  
    The distribution of magnetic-to-radiation pressure ratios appears relatively flat as a function of star formation rate surface density.}
    \label{fig:urat-sigsfr}
\end{figure}

\section{Results}\label{sec:results}

In the following section, the results from the pressure estimates described above (i.e., thermal gas, radiation, and magnetic/CR) are presented.    
This is first done by comparing each of these locally-measured internal pressure terms with one another, followed by looking at how their total sum compares to the required pressure for the interstellar gas to stay in equilibrium within the gravitational potential of a galaxy on kpc scales.  
Estimated pressures and other physical quantities for each region are provided in Table~\ref{tbl:pressures} of the Appendix.  


\subsection{Individual Pressure Terms}\label{sec:results-int}
In Figure \ref{fig:urat-th-sigsfr}, thermal gas-to-radiation pressure ratios are plotted as a function of $\Sigma_{\rm SFR}$.  
Extranuclear and nuclear regions from the SFRS and GOALS galaxies are identified.  
There are two sources with thermal gas-to-radiation pressure ratios $\gtrsim$5.  
One is the nucleus of NGC\,4594, which is known to host an active galactic nuclei (AGN) \citep{jm10} that is most likely driving this large ratio due to a significant amount of radio emission associated with accretion relative to the  thermal dust emission.  
The second is NGC\,2403\,Enuc.\,1\,a, which has a counterpart in the Open Supernova Catalog \citep[OSC;][]{OSC17}, but was not categorized by \citet{stl20} as a potential supernova remnant (SNR) due to the lack of an optical counterpart.  
NGC\,2403\,Enuc.\,1\,A was previously identified as a possible radio supernova candidate by \citet[TH-2;][]{th94} and was also detected in X-rays by \citet[SP-12;][]{sp03}, thus indeed appearing consistent with being an optically-faint SNR. 

From Figure \ref{fig:urat-th-sigsfr}, the radiation pressure is found to be a factor of $\sim$2.2 larger, on average when taking a median, compared to the thermal gas pressure among all regions. 
Also, while there appears to be a clear trend of decreasing thermal gas-to-radiation pressure ratios with increasing values of  $\Sigma_{\rm SFR}$, a least-squares fit to the data (dot-dashed line) indicates that the decrease is only marginally steeper than what is expected given the correlated nature of the axes (i.e., thin solid line).  
This result suggests that the radiation pressure (i.e., infrared dust emission) does not increase that much more rapidly with star formation activity than the thermal gas pressure (i.e., free-free emission). 

In Figure \ref{fig:apD-th}, thermal gas-to-radiation pressure ratios are plotted as a function of aperture diameter in linear units. 
Various subsets of the sample are identified using different plotting symbols.  
Additionally, running medians for all regions, as well as all nuclear and extranuclear regions, are shown by solid lines using large filled circles, triangles, and squares, respectively.  
What is clearly shown is that the ratio of thermal gas-to-radiation pressures decrease with increasing physical size of each region, likely indicating that the thermal pressure (i.e., free-free emission) is more centrally peaked on individual star-forming regions than the radiation pressure (i.e., infrared dust emission).

In Figure \ref{fig:urat-sigsfr}, magnetic-to-radiation pressure ratios are instead plotted as a function of $\Sigma_{\rm SFR}$.  
Again, extranuclear and nuclear regions from the SFRS and GOALS galaxies are identified.  
Overall, the distribution of magnetic-to-radiation pressure ratios appears to be relatively flat as a function of $\Sigma_{\rm SFR}$. 
Given that the magnetic pressure estimates were made under the minimum energy assumption, any comparison with the CR and radiation pressures will yield similar results, and therefore are not shown.

There are two sources that have magnetic-to-radiation pressure ratios $\gtrsim$5: NGC\,5866 and NGC\,5194\,Enuc.\,11\,C.     
NGC\,5866 is an S0 galaxy that hosts a known AGN \citep{jm10}, which is most likely driving this large ratio.  
NGC\,5194\,Enuc.\,11\,C has a counterpart in the OSC \citep{OSC17}, but was not categorized by \citet{stl20} as a potential SNR due to the lack of an optical or X-ray counterpart.  
However, as pointed out by \citet{maddox07}, this source may in fact be an optically-faint SNR.  

\begin{figure}
\epsscale{1.25}
    \centering
    \plotone{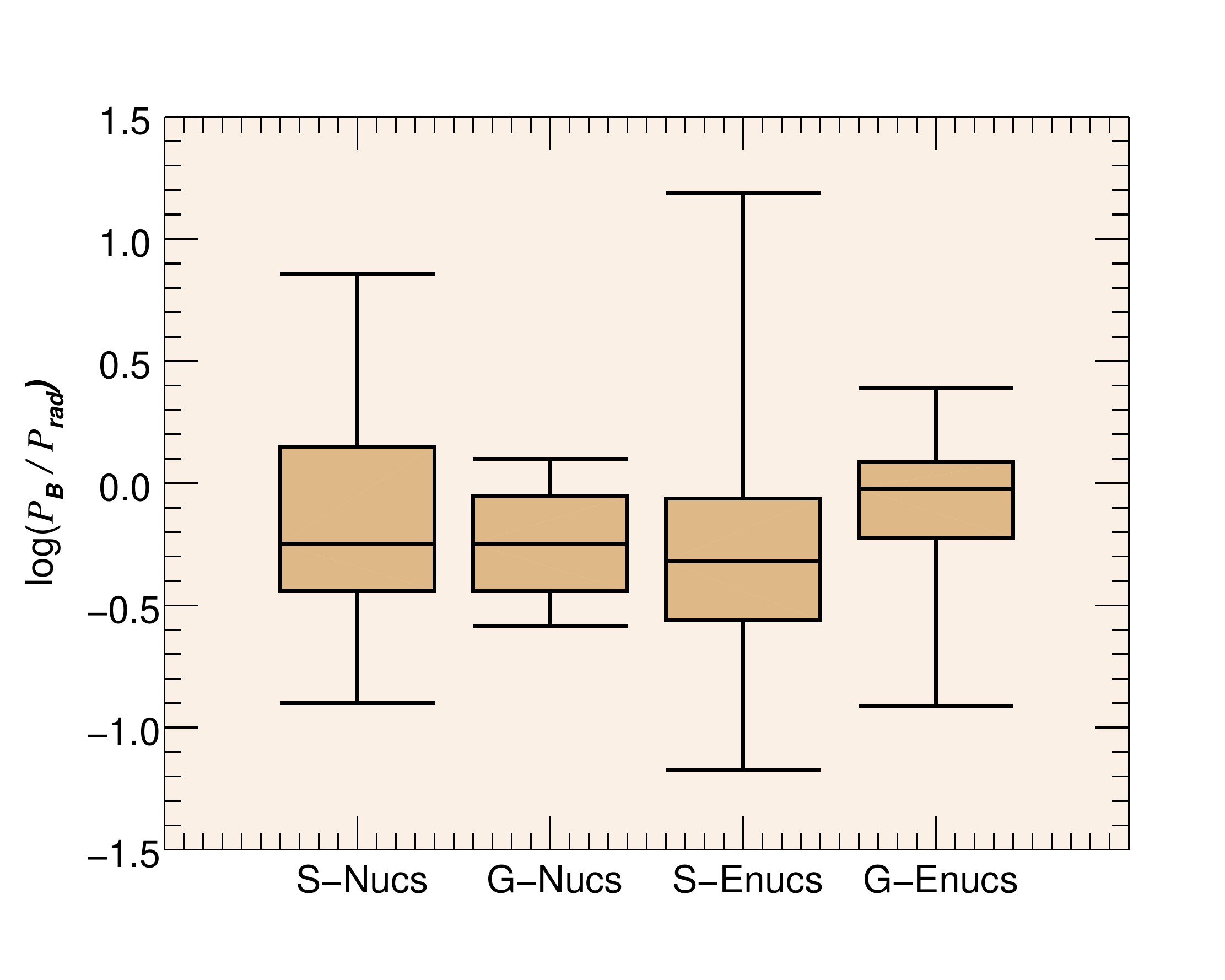}
    \caption{A box and whiskers plot illustrating the minimum, lower quartile, median, upper quartile, and maximum values for the logarithmic ratio of magnetic-to-radiation pressures for the SFRS and GOALS nuclear and extranuclear regions.
    The median ratios are quite similar for the nuclear regions in both the SFRS and GOALS galaxies, however, the ratio for the extranuclear regions in the GOALS galaxies is a factor of $\sim$2 larger than that for the SFRS extranuclear regions.  }
    \label{fig:boxplot}
\end{figure}

In Figure \ref{fig:boxplot}, a box and whiskers plot is shown indicating the minimum, lower quartile, median, upper quartile, and  maximum values for the ratio of magnetic-to-radiation pressures for the SFRS and GOALS nuclear and extranuclear regions.
This is done to see if there is any significant differences in the pressure ratios among the galaxy samples and nuclear and extranuclear environments. 

\begin{figure}
\epsscale{1.25}
    \centering
    \plotone{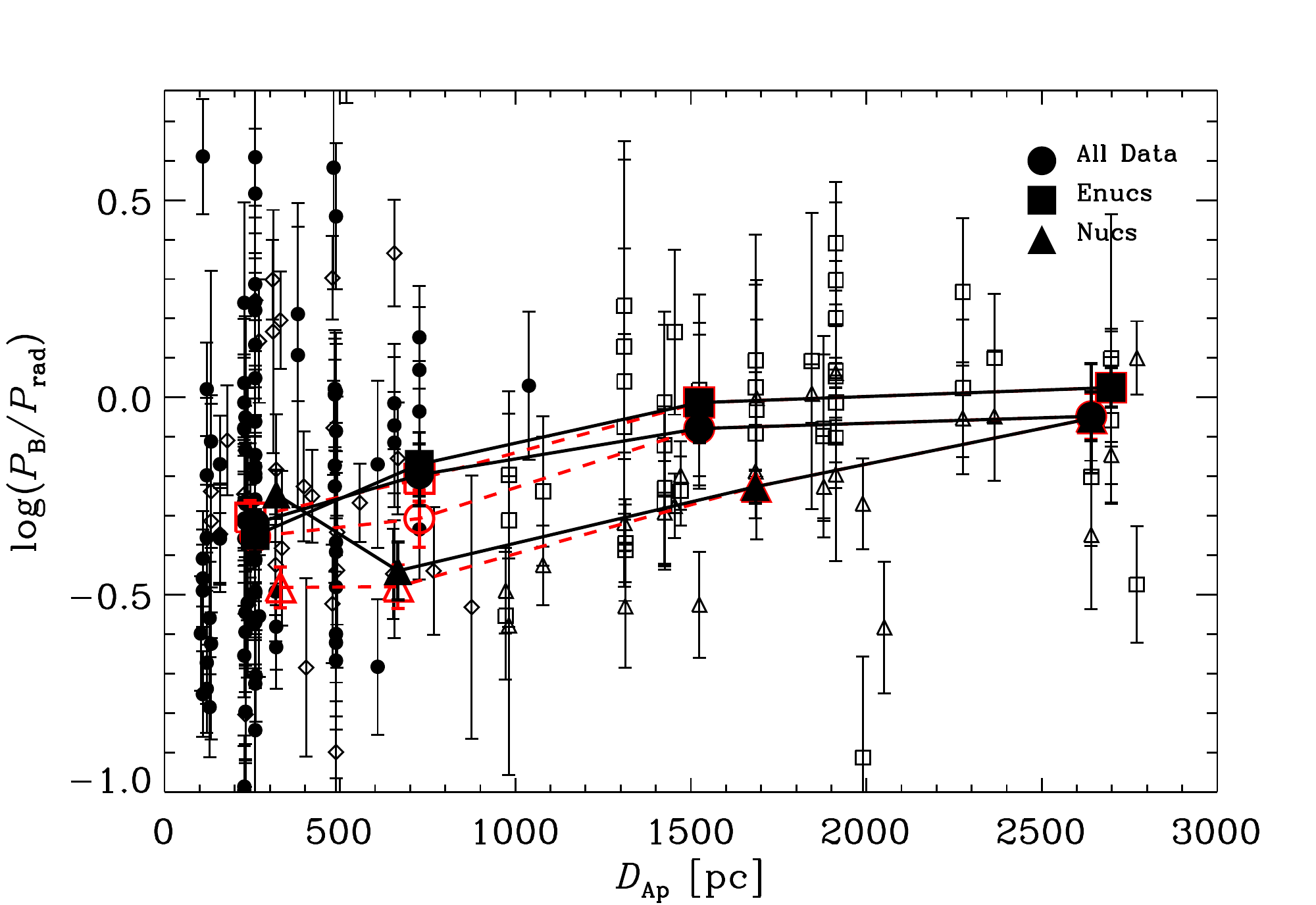}
    \caption{Logarithmic magnetic-to-radiation pressure ratio plotted as a function of aperture diameter in physical units.   
    Subsets of the sample are identified using different plotting symbols: SFRS nuclear regions (open diamonds); SFRS extranuclear regions (filled circles); GOALS nuclear regions (open triangles); and GOALS extranuclear regions (open squares).  
    Running medians for all regions, as well as all nuclear and extranuclear regions, are shown by solid lines using large filled circles, triangles, and squares, respectively.  
    The red dashed-line and open symbols indicate how the magnetic pressure estimates change by letting the path length of the synchrotron emitting medium scale with the physical diameter of the photometric apertures for extranuclear regions resolved on scales $<$1\,kpc after subtracting a local background.  
    The magnetic-to-radiation pressure ratios appear to be sensitive to the physical size of each region, showing a steady increase with increasing physical size of the aperture.    }
    \label{fig:apD}
\end{figure}

The median ratios appear to be quite similar for the nuclear regions in both the SFRS and GOALS galaxies. 
However, there does appear to be a measurable difference in the pressure ratios for the extranculear regions.  
The magnetic-to-radiation pressure ratio for the extragalactic regions in the GOALS galaxies is a factor of $\sim$2 larger than that for the SFRS extranuclear regions. 

Given that the GOALS galaxies are at much larger distances than the SFRS galaxy sample, the physical sizes of the regions being studied in the GOALS galaxies are larger by a factor of $\sim$6.5, on average (i.e., a median aperture diameter of 1.7\,kpc for the GOALS galaxies compared to a median aperture diameter of 260\,pc for the SFRS galaxies).  
To see how this affects the magnetic-to-radiation pressure ratios, they are plotted as a function of aperture diameter in linear units in Figure \ref{fig:apD}. 

In Figure \ref{fig:apD}, subsets of the sample are identified using different plotting symbols, which include the SFRS nuclear regions, SFRS extranuclear regions, GOALS nuclear regions, and GOALS extranuclear regions.  
Running medians for all regions, along with the subset of nuclear and extranuclear regions, are also shown by solid lines indicating an increase in the magnetic-to-radiation pressure ratios with increasing physical size of the apertures.  
Red dashed-lines indicate how the the running medians change when using magnetic pressure estimates that let the path length of the synchrotron emitting medium scale with the physical diameter of the photometric apertures for extranuclear regions resolved on scales $<$1\,kpc after subtracting a local background (see \S\ref{sec:PB}).    
As is clearly shown, such a change does not significantly change the magnetic-to-radiation pressure ratios for such regions.  
Consequently, the magnetic-to-radiation pressure ratios appear to be sensitive to the physical size of each region, showing a steady increase with increasing physical size of the aperture.


\begin{figure}
\epsscale{1.25}
    \centering
    \plotone{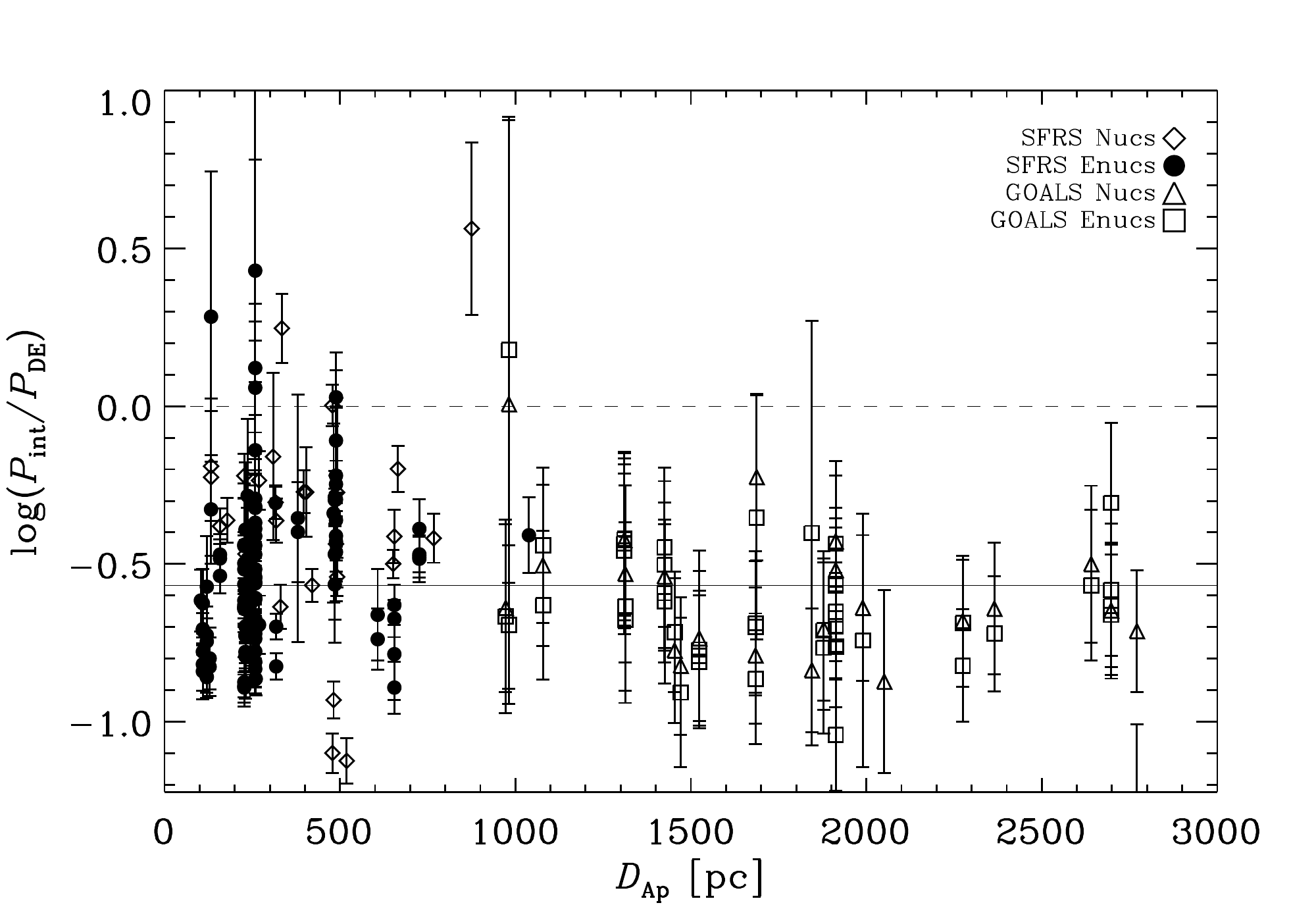}
    \caption{Logarithmic ratios of the total internal-to-dynamical equilibrium pressures plotted as a function of aperture diameter in physical units.  
    The dynamical equilibrium pressures are estimated via a linear interpolation using the measured star formation rate surface densities and the correlation between $\Sigma_{\rm SFR}$ and $P_{\rm DE}$ found by \citet{sun20}. 
    Subsets of the sample are identified using different plotting symbols.  
    The dashed line indicates a ratio of unity, while the solid line illustrated the median ratio.  
    Neglecting any additional pressure term associated with turbulent motions of the gas, only $\approx$5\% of the star-forming regions appear to be over-pressurized relative to the dynamical equilibrium pressure, each of which are measured on $\lesssim$1\,kpc scales.  
    }
    \label{fig:apD-tot}
\end{figure}

\subsection{Total Internal Pressure vs. Dynamical Equilibrium Pressure}
\label{sec:ptot}
Having estimates for a number of internal pressure terms, their net effect can be used to determine what fraction of sources appear to be over-pressurized relative to what is required for the interstellar gas to stay in equilibrium within the gravitational potential of a galaxy.  
Unfortunately, the current analysis lacks an independent estimate of the dynamical equilibrium pressure to actually determine what fraction of sources may be over-pressurized.  

The dynamical equilibrium pressure typically includes a term to account for the weight associated with self-gravity of the ISM disk \citep{spitz42, elmegreen89} and another term to account for the weight of the ISM due to stellar gravity in the potential well of the stars \citep{spitz42, br04}.  
Following this approach, the dynamical equilibrium pressure has recently been expressed by \citet{sun20} such that
\begin{equation}
    P_{\rm DE} = \frac{\pi G}{2} \Sigma_{\rm gas}^{2} + \Sigma_{\rm gas}\sqrt{2G\rho_{*}}\sigma_{\rm gas, z}, 
    \label{eq:Pdyn}
\end{equation}
where, $\Sigma_{\rm gas} = \Sigma_{\rm mol} + \Sigma_{\rm atom}$ is the total gas surface density, $\rho_{*}$ is the stellar mass volume density near the disk mid-plane, and $\sigma_{\rm gas,z}$ is the velocity dispersion of the gas perpendicular to the disk.  
Since such independent estimates of $P_{\rm DE}$ are not available for the regions being studied here, we assign values based on the work of \citet{sun20}, who present kpc-scale dynamical equilibrium pressures for 1762 regions in 28 nearby galaxies.  

Total internal pressures are calculated as $P_{\rm int} = P_{\rm th} + P_{\rm rad} + P_{\rm CR} + P_{\rm B}$ and are compared with the dynamical equilibrium pressures estimated via linear interpolation using the measured star formation rate surface densities and the correlation between $\Sigma_{\rm SFR}$ and $P_{\rm DE}$ found by \citet{sun20}, which has a scatter of $\approx$0.2\,dex.    
While this correlation was strictly established on kpc-scale measurements of $\Sigma_{\rm SFR}$ and $P_{\rm DE}$, it should still provide a qualitatively meaningful comparison for the present investigation even though $\Sigma_{\rm SFR}$ is measured on scales spanning $\sim100$\,pc to $\sim$3\,kpc.  
For reference, the median dynamical equilibrium pressures for the SFRS and GOALS nuclear and extranuclear regions are: $\langle P_{\rm DE}\rangle = (188 \pm 53)\times10^{-12}$, $(143 \pm 15)\times10^{-12}$, $(1773 \pm 287)\times10^{-12}$, and $(466 \pm 72)\times10^{-12}\,{\rm dyn\,cm^{-2}}$, respectively.  


Using this relation to estimate the dynamical equilibrium pressures for each region, the median ratio of $P_{\rm int}/P_{\rm DE}$ is found to be $\approx$0.3, with very few sources (i.e., $\approx$5\%) appearing to be over-pressurized.  
In Figure \ref{fig:apD-tot}, the ratio of the total internal-to-dynamical equilibrium pressures are plotted as a function of aperture diameter.  
Of the sources that are found to be over-pressurized, each of them were measured in sub-kpc apertures, again suggesting that some combination of the the internal pressure terms are likely stronger on smaller spatial scales.  

It is worth noting that there is a missing internal pressure term in this analysis associated with the turbulent motions in the gas. 
However, given that \citet{sun20} report that the turbulent gas pressure both correlates with, and almost always exceeds, the dynamical equilibrium pressure on kpc scales, its inclusion in the calculation above will simply result in every region being over-pressurized.  
Again, taking the results from \citet{sun20}, who measures turbulent gas pressures on 120\,pc scales, and linearly interpolating from our estimates of the star formation rate surface densities, the median turbulent gas pressures for the SFRS and GOALS nuclear and extranuclear regions are: $\langle P_{\rm turb}\rangle = (200 \pm 57)\times10^{-12}$, $(152 \pm 16)\times10^{-12}$, $(1874 \pm 304)\times10^{-12}$, and $(494 \pm 76)\times10^{-12}\,{\rm dyn\,cm^{-2}}$, respectively. 
Indeed, including the estimates of $P_{\rm turb}$ in the calculation of $P_{\rm int}$ results in every region being marginally over-pressurized, with a median ratio of $P_{\rm int}/P_{\rm DE} \approx 1.3$.   

The above result arises from that fact that in feedback models, expanding H{\sc ii} regions, radiation pressure, and supernovae all drive the observed turbulence, which in turn supports the dynamical equilibrium pressure of the ISM \citep[e.g.,][]{tqm05, ostriker-shetty11, faucher13, ostriker-kim22}.   
Consequently, in this situation, it is not appropriate to add the stellar feedback pressures with the turbulent pressure. 
Though, this picture may be slightly more complicated given that recent models by \citet{krumholz18} argue that gas accretion should be included as an additional mechanism driving turbulence.

\section{Discussion and Conclusions}\label{sec:disc}
Observational studies, either for entire galaxies or smaller regions within them defined by a specific volume or density threshold, find that the gas depletion time $t_{\rm dep} = \Sigma_{\rm gas}/\Sigma_{\rm SFR}$, where $\Sigma_{\rm gas}$ is the gas surface density, is always $\sim 1-3$ orders of magnitude longer than the free-fall time $t_{\rm ff} \propto 1/\sqrt{G\rho}$, where $\rho$ is the volume density of the gas \citep[e.g.,][]{kt07,evans09,krumholz12,federrath13}.  
In contrast, numerical simulations of star cluster formation that do not include any form of feedback generally produce $t_{\rm dep} \sim t_{\rm ff}$ \citep[e.g.,][]{klessen00,klessen01,bate03,bonnel03}, clearly indicating its importance in regulating the star formation process.  
In the present analysis, a number of primary candidates for regulating star formation (i.e., thermal gas, radiation, and magnetic pressure estimates) are compared as a function of galaxy environment, star formation activity, and physical areas over which the quantities are measured.  


In comparing the thermal gas and radiation pressure estimates presented here with those in the literature, they appear highly consistent to values measured via other methods.  
For instance, \citet{barnes21} recently looked into the various stellar (pre-supernova) feedback mechanisms (i.e., photoionized gas, radiation, winds) from $\approx$6000 H{\sc ii} regions within a sample of 19 nearby galaxies on $\sim 50-100$\,pc scales included in Physics at High Angular resolution in Nearby GalaxieS (PHANGS) sample.  

In their analysis, they calculated upper and lower limits for each of the pressure estimates using minimum (i.e., clumpy structure: $\sim$10\,pc) and maximum (i.e., smooth structure: $\sim$100\,pc) sizes for each region, respectively.
Thermal gas pressure estimates were made by using the [S{\sc ii}] $\lambda\lambda$6716,31 doublet 
to obtain electron densities and additionally assuming a fixed electron temperature of 8000\,K.  
Radiation pressures were estimated by converting H$\alpha$ measurements into bolometric luminosities.  
The mean values of their lower and upper limit pressure estimates for their entire sample are shown as vertical lines in Figures \ref{fig:Praddist} and \ref{fig:Pthdist}, which clearly illustrate that our radiation and thermal-gas pressure estimates are largely consistent with theirs.  

Similarly, minimum-energy magnetic field strengths, and therefore pressures, are also consistent with other estimates in the literature.  
Typical values of the isotropic turbulent field strengths in star-forming galaxy arms and bars can range between $20-30\,\mu$G, while central starbursting regions have field strengths as high as $50-100\,\mu$G \citep[e.g.,][and reference therein]{beck15}.  
As shown in Figure~\ref{fig:Bmindist}, these values are in good agreement with the minimum-energy magnetic field strength distributions for extranuclar and nuclear regions, as well as for normal galaxies and the starbursting GOALS galaxies.

The results from this analysis appear to indicate that the physical area, and not necessarily environmental location (e.g., nuclear vs. extranuclear regions) is the dominant factor in determining the relative importance of individual pressure terms around star-forming regions.  
For instance, the thermal gas pressure appears to be similar to, and in some cases exceeds, the radiation pressure on sub-kpc scales.
On the other hand, it seems that the magnetic pressure is typically sub-dominant on sub-kpc scales and only starts to play a significant role on few-kpc scales relative to radiation pressure.  

The result of finding the magnetic pressure to start to dominate over the radiation pressure on galaxy-disk scales is consistent with expectations based on the well-known far-infrared--radio correlation for galaxies \citep[e.g., ][]{de85,gxh85,yrc01}, whereby the ratio of magnetic-to-radiation energy densities must be $\gtrsim$1 to ensure that synchrotron and inverse Compton losses to radio continuum emitting cosmic-ray electrons is fixed \citep[e.g.,][]{jc92,ejm09c}.  
It is also worth pointing out that even if the minimum-energy magnetic field is significantly underestimating the true magnetic field strength on small scales, as has been suggested for starburst galaxies \citep[e.g.,][]{tt06} and their central molecular zones \citep[e.g.,][]{Yoast-Hull16}, such that it is similar to what is required for equipartition with the hydrostatic pressure of the molecular clouds, this should not affect our overall results.  
Observations suggest that molecular clouds are magnetically supercritical by a factor of $\sim$2 \citep[][and references therein]{crutcher12} while simulations also indicate that a magnetic field of this strength is only likely to suppress the overall star formation activity on such scales by factor of a few compared to the purely hydrodynamic case \citep{pb09,padoan11}.  

These results likely suggest that it is the combination of all feedback mechanisms that work together to regulate star formation on a range of physical scales, keeping the process inefficient both in individual clouds and galaxy disks.  
This conclusion is also supported by the relatively constant ratio of total internal-to-dynamical equilibrium pressures as a function of physical aperture area. 
These findings are in agreement with \citet{krumholz14}, who suggested that magnetic fields alone are unlikely to be the single driver that is able to keep star formation rates regulated, but rather a combination of magnetic fields with stellar feedback as the prime candidate that regulate star formation. 
Accordingly, these observations support a picture where stellar feedback processes in mature galaxy disks (e.g., thermal gas and radiation pressure on dust) are able to be highly effective on the scales of individual star-forming complexes.  
However, in the transition to larger (i.e., few-kpc) scales, other processes (e.g., a galaxy's large-scale magnetic field) become the dominant mechanism regulating star formation.  
This scenario is consistent with the magnetohydrodynamical calculations of \citet{pb09}, who found that magnetic fields primarily provide support on large scales to low-density gas, whereas radiation is found to strongly suppress small-scale fragmentation of clouds by increasing the temperature in the high-density material near the protostars.

While only a handful of stellar feedback process have been looked at in this study, which is additionally limited by the fixed angular resolution of the data in hand, the distance of the sample galaxies, and the lack of independent estimates for the dynamical equilibrium pressure,
the general conclusion of the importance of multiple processes being effective on different physical scales to collectively regulate the star formation efficiency of galaxies is most likely robust.  
In considering the effects of wind pressure in addition to the pressure terms from stellar feedback also considered here, \citet{barnes21} found that each of these processes were generally comparable, and their sum was in excess of the kpc-scale dynamical equilibrium pressure for 99\% of their sample when a clumpy ($\sim$10\,pc) structure was assumed.  
When a smooth ($\sim$100\,pc) structure was assumed, the fraction of sources with a total internal pressure larger than the kpc-scale dynamical equilibrium pressure decreases.  
This is in general agreement with the results of this analysis, which is limited to $\gtrsim$100\,pc scales.  
Consequently, it appears that each of these physical feedback processes exhibit differing degrees of relative importance on sub-kpc scales, and that their total sum is less than the dynamical equilibrium pressure on kpc scales unless an additional contribution from the turbulent gas pressure is included.  
However, to definitively determine the exact pressure balance among each of these processes, and properly assess the physical scale(s) over which each term may dominate, requires substantially more data across the electromagnetic spectrum and at a range of angular resolutions.  




\begin{acknowledgments}
EJM would like the thank the anonymous referee for a very careful reading of the text and substantial comments that helped improve the content of this paper.  
EJM would like to thank Sean Linden for making all of his catalogs available for this study.  
EJM is grateful to Shane Davis for a close reading of the text and for a number of very useful comments that helped improved the paper.  
EJM is also grateful to Rainer Beck for a number of stimulating discussions that improved this work.  
The National Radio Astronomy Observatory is a facility of the National Science Foundation operated under cooperative agreement by Associated Universities, Inc. 
\end{acknowledgments}

\facilities{VLA, Spitzer, GALEX}

\bibliography{master_ref}{}
\bibliographystyle{aasjournal}

\appendix
\label{sec:appendix}
In Table~\ref{tbl:pressures}, estimated physical quantities for all regions are provided.  

\begin{longrotatetable}
\begin{deluxetable}{lcc|cc|cccc|cc}
\tablecolumns{11}
\tablecaption{Estimated Physical Quantities for Each Region \label{tbl:pressures}}
\tabletypesize{\scriptsize}
\tablewidth{0pt}
\tablehead{
\colhead{Source ID} & \colhead{R.A.} & \colhead{Decl.} & \colhead{$D_\mathrm{Ap}$} & \colhead{$\Sigma_\mathrm{SFR}$} & \colhead{$P_\mathrm{B}$} & \colhead{$P_\mathrm{CR}$} & \colhead{$P_\mathrm{rad}$} & \colhead{$P_\mathrm{th}$} & \colhead{$P_\mathrm{DE}^{\dagger}$} & \colhead{$P_\mathrm{turb}^{\dagger}$}\\
\colhead{} & \colhead{(J2000)} & \colhead{(J2000)} & \colhead{(kpc)} & \colhead{($M_{\odot}$\,yr$^{-1}$\,kpc$^{-2}$)} & \multicolumn{6}{c}{(10$^{-12}$\,dyn\,cm$^{-2}$)}
}
\startdata
\cutinhead{SFRS Nuclear Regions}
                NGC\,0855   &           $02~14~03.68$  &           $+27~52~37.9$  &0.330  & $  0.29\pm  0.02$  & $ 13.94\pm  3.90$  & $  5.08\pm  1.42$  & $  8.89\pm  0.44$  & $ 10.76\pm  3.18$  &$  167.18\pm  14.60$  &$  178.22\pm  15.57$  \\
                NGC\,0925   &           $02~27~17.00$  &           $+33~34~43.0$  &0.310  & $  0.04\pm  0.02$  & $  4.55\pm  3.20$  & $  1.66\pm  1.17$  & $  3.10\pm  0.31$  & $  3.95\pm  2.82$  &$   19.15\pm   9.77$  &$   21.84\pm  11.14$  \\
                NGC\,1097   &           $02~46~18.98$  &           $-30~16~28.8$  &0.482  & $  2.50\pm  0.20$  & $ 57.60\pm 15.60$  & $ 20.98\pm  5.68$  & $ 68.83\pm  3.39$  & $ 26.36\pm  7.55$  &$ 1481.86\pm 121.42$  &$ 1567.14\pm 128.41$  \\
                NGC\,1482   &           $03~54~38.97$  &           $-20~30~07.8$  &0.767  & $  3.40\pm  0.51$  & $180.71\pm 66.84$  & $ 65.83\pm 24.35$  & $497.38\pm 24.86$  & $ 24.35\pm  9.40$  &$ 2013.84\pm 299.83$  &$ 2129.16\pm 317.00$  \\
                NGC\,2798   &           $09~17~22.85$  &           $+42~00~00.4$  &0.876  & $  0.28\pm  0.17$  & $124.44\pm 95.67$  & $ 45.33\pm 34.85$  & $423.48\pm 21.08$  & $  6.55\pm  5.09$  &$  164.28\pm  99.28$  &$  175.17\pm 105.86$  \\
                NGC\,2841   &           $09~22~02.67$  &           $+50~58~35.7$  &0.479  & $  0.84\pm  0.05$  & $ 13.08\pm  3.11$  & $  4.76\pm  1.13$  & $  6.51\pm  0.34$  & $ 15.37\pm  3.89$  &$  498.74\pm  31.93$  &$  528.51\pm  33.84$  \\
                NGC\,3049   &           $09~54~49.56$  &           $+09~16~16.1$  &0.652  & $  0.66\pm  0.05$  & $ 26.74\pm  6.92$  & $  9.74\pm  2.52$  & $ 74.90\pm  3.58$  & $ 11.63\pm  3.19$  &$  388.72\pm  29.19$  &$  412.27\pm  30.96$  \\
                NGC\,3077   &           $10~03~19.15$  &           $+68~43~59.9$  &0.130  & $  2.50\pm  0.16$  & $ 37.29\pm  8.96$  & $ 13.58\pm  3.27$  &           \nodata  & $ 50.71\pm 12.95$  &$ 1479.37\pm  96.52$  &$ 1564.51\pm 102.07$  \\
\enddata
\tablecomments{$^{\dagger}$\,Pressure estimates derived using empirical relations between $P_{\rm DE}$, $P_{\rm turb}$ and $\Sigma_{\rm SFR}$ given in \citet{sun20}.  Associated uncertainties provided here do not include the intrinsic scatter of these empirical relations.}
\end{deluxetable}
\end{longrotatetable}



\end{document}